\documentclass[acmlarge]{acmart}

\usepackage{booktabs} % For formal tables
\usepackage[linguistics]{forest}
\usepackage{tikz}
\usetikzlibrary{cd}
\usepackage{xcolor}
\usepackage[labelsep=quad,indention=10pt]{subfig}
\usepackage{mathpartir} % Inference rules
\usepackage[ruled]{algorithm2e} % For algorithms
\usepackage{setspace}

\SetAlFnt{\small}
\SetAlCapFnt{\small}
\SetAlCapNameFnt{\small}
\SetAlCapHSkip{0pt}
\IncMargin{-\parindent}

\usepackage{amsmath}
\usepackage{listings}
\usepackage{racketlst}
\usepackage{xspace}
\usepackage{letltxmacro}
\newcommand*{\SavedLstInline}{}
\LetLtxMacro\SavedLstInline\lstinline
\DeclareRobustCommand*{\lstinline}{%
  \ifmmode
    \let\SavedBGroup\bgroup
    \def\bgroup{%
      \let\bgroup\SavedBGroup
      \hbox\bgroup
    }%
  \fi
  \SavedLstInline
}

\definecolor{colorMATH}{HTML}{3E5BA9}
\definecolor{colorFAC}{HTML}{9F1D20}

\newcommand{\blue}[1] {{\color{colorMATH} #1}}
\newcommand{\fcol}[1] {{\color{colorFAC} #1}}
\newcommand{\bmth}[1] {{\color{colorMATH} $#1$}}
\newcommand{\rmth}[1] {{\color{colorFAC} $#1$}}
\newcommand{\code}[1]{\lstinline{#1}}

\newcommand{\lang}[0]{$\lambda_\textsc{FE}$\xspace}
\newcommand{\racets}[0]{$\textsc{Racets}$\xspace}
\newcommand{\facet}[3]{{\fcol{\ensuremath{\langle \, {#1} \: ? \: {#2} \diamond \: {#3} \, \rangle}}}}
\newcommand{\formfacet}[3]{{\fcol{\ensuremath{\langle \langle \, {#1} \: ? \: {#2} \diamond \: {#3} \,  \rangle \rangle}}}}
\newcommand{\var}[1]{\mathit{#1}}

\begin{document}

\title{\racets: Faceted Execution in Racket}

\author{Kristopher Micinski}
\affiliation{%
  \institution{Haverford College}
  \city{Haverford}
  \state{PA}
  \postcode{19041}
  \country{USA}}
\email{kris@cs.haverford.edu}
\author{Zhanpeng Wang}
\affiliation{%
  \institution{Haverford College}
  \city{Haverford}
  \state{PA}
  \postcode{19041}
  \country{USA}
}
\email{zwang10@haverford.edu}
\author{Thomas Gilray}
\affiliation{%
 \institution{University of Alabama, Birmingham}
 \city{Birmingham}
 \state{AL}
 \country{USA}}
\email{gilray@uab.edu}

\begin{abstract}
Faceted Execution is a linguistic paradigm for dynamic
information-flow control. Under faceted execution, secure program data
is represented by \emph{faceted} values: decision trees that encode how the
data should appear to its owner (represented by a label) versus everyone
else. When labels are allowed to be first-class (i.e., predicates that
decide at runtime which data to reveal), faceted execution enables
\emph{policy-agnostic programming}: a programming style that allows
privacy policies for data to be enforced independently of code that
computes on that data.

To date, implementations of faceted execution are relatively
heavyweight: requiring either changing the language runtime or the
application code (e.g., by using monads). Following Racket's
languages-as-libraries approach, we present Racets: an implementation
of faceted execution as a library of macros.  Given Racket's
highly-expressive macro system, our implementation follows relatively
directly from the semantics of faceted execution. To demonstrate how
\racets can be used for policy-agnostic programming, we use it to
build a web-based game of Battleship. Our implementation sheds light
on several interesting issues in interacting with code written without
faceted execution. Our \racets implementation is open source, under
development, and available online.
\end{abstract}

%% \begin{CCSXML}
%% <ccs2012>
%%  <concept>
%%   <concept_id>10010520.10010553.10010562</concept_id>
%%   <concept_desc>Computer systems organization~Embedded systems</concept_desc>
%%   <concept_significance>500</concept_significance>
%%  </concept>
%%  <concept>
%%   <concept_id>10010520.10010575.10010755</concept_id>
%%   <concept_desc>Computer systems organization~Redundancy</concept_desc>
%%   <concept_significance>300</concept_significance>
%%  </concept>
%%  <concept>
%%   <concept_id>10010520.10010553.10010554</concept_id>
%%   <concept_desc>Computer systems organization~Robotics</concept_desc>
%%   <concept_significance>100</concept_significance>
%%  </concept>
%%  <concept>
%%   <concept_id>10003033.10003083.10003095</concept_id>
%%   <concept_desc>Networks~Network reliability</concept_desc>
%%   <concept_significance>100</concept_significance>
%%  </concept>
%% </ccs2012>
%% \end{CCSXML}

%% \ccsdesc[500]{Computer systems organization~Embedded systems}
%% \ccsdesc[300]{Computer systems organization~Redundancy}
%% \ccsdesc{Computer systems organization~Robotics}
%% \ccsdesc[100]{Networks~Network reliability}


\maketitle

\renewcommand{\shortauthors}{Micinski, Wang, and Gilray}

\section{Introduction}

As information systems become more interconnected and complex, they consume an
ever-growing amount of private data. System designers communicate to users
how their data may be used via a \emph{privacy policy}. Unfortunately,
implementing such policies correctly is challenging: users often have
partial control over the policy (e.g., whether their phone number is
publicly visible or private) and policies can change frequently.
Not only can specific privacy policies be highly dynamic (dependent on runtime values),
but the process of improving privacy policies can be highly dynamic across time.
As policies evolve, developers face massive (re)engineering efforts to
ensure that implementations continue to match the policy at every relevant
point in the codebase.

\emph{Policy-agnostic programming} is a linguistic paradigm that
decouples the implementation of privacy policies from the code that
operates on sensitive data. This frees developers to write programs mostly
as they would for insecure code, without inserting specific logic to manage
information-flow policies directly into application code. Instead, data is
labeled with its policy as it enters the system and such labels propagate through
the program, alongside data, as computation progresses. When a secure value needs
to be introspected upon (or propagates outside the application), its policy
can be invoked at this point dynamically. This paradigm aims to permit
code manipulating sensitive data to be written in a manner entirely orthogonal to
policies themselves.

\emph{Faceted execution} (FE) is a highly expressive 
language semantics enabling policy-agnostic programming~\cite{Austin:2012}. In FE, dynamic information-flow
monitors instrument the program, encoding sensitive values as \emph{faceted values}:
decision trees specifying different views of data according to different possible security labels.
For example, the faceted value \facet{Alice}{\code{true}}{\code{false}}
represents a value that should appear to Alice as \blue{\code{true}} and to
everyone except Alice as \blue{\code{false}}. Faceted execution propagates distinct facets
of a value by extending core linguistic primitives (such as function application).
For example, consider the application \blue{\code{(x true)}} where
\code{x} is \blue{\facet{Alice}{\lambda x.~\code{true}}{\code{not}}}.
Racket's standard function application will fail here because Racket's
\blue{\code{ppapp}} expects a procedure rather than a facet. Instead, the proper way
to interpret function application on faceted values is to distribute the
application over all (in this case, both) facets, producing
\facet{Alice}{\code{true}}{\code{false}}: if Alice is viewing, the application yielded true,
otherwise it yielded false, so both are computed until the value is explicitly observed
with specified permissions. Many other core forms (such
as \code{if}, \code{set!}, etc...) require similar changes to handle faceted
values correctly.

Policy-agnostic programming promotes the idea that programmers should
be able to write programs ``normally'', without concerning themselves
with how privacy policies are enforced. Unfortunately, the
relatively foundational linguistic changes required to enable faceted
execution have hindered implementations thusfar. Dynamic generation
of first-class security labels, tracking an arbitrary number of
facets per value, and keeping faceted-value trees in a canonical order,
are all central challenges in any practical implementation. For example, the
first implementation of FE (by \citet{Austin:2012}) extended
a JavaScript interpreter to account for faceted values. Other
implementations use monads~\cite{Schmitz:2016}
or rely upon third-party macro systems~\cite{Yang:2016}. We
know of no existing implementation of FE that aims to interoperate seamlessly
with code written in the host language. By contrast, Scheme boasts a powerful
hygienic macro system that allows essentially any linguistic form
to be modified arbitrarily.

In this paper we present \racets, an implementation of policy-agnostic
programming in Racket via macros~\cite{Kohlbecker:1986, Dybvig:1992}. \racets provides facilities for
creating policies and faceting secure data with those
policies. \racets also extends several core forms in Racket to work
with faceted values (our implementation is detailed in
Section~\ref{sec:implementation}). We have used \racets to implement a
small server-based board-game (detailed in Sections~\ref{sec:overview}
and~\ref{sec:implementation}). Relevant related work is presented in
section~\ref{sec:related}. We see \racets as a promising prototype for
policy-agnostic programming in Racket, and conclude with discussion of
future directions in Section~\ref{sec:conclusion}.

\section{Overview of Faceted Execution}
\label{sec:overview}

To introduce faceted execution more concretely, we present the implementation
of Battleship, a small guessing game, in \racets (this section presents a
distilled version of our case study in
Section~\ref{sec:implementation}). In this game each player has a
private board of coordinates, at which they place ``ships''. The players
hide their boards from each other as play progresses in rounds. Each
turn a player guesses the position of a ship on the other player's
board. If the guess is successful the tile is removed from the
board and a hit is declared publicly. Play ends once one player's board
has no remaining tiles, at which point that player loses.

We implement game boards as lists of cons cells representing the
\blue{$(x,y)$} coordinates of ships. Board creation simply returns an empty
list, and adding a piece is done via \lstinline|cons|:

\begin{lstlisting}[language=Racket,escapechar=|,name=example]
(define (makeboard) '())
(define (add-piece board x y) (cons (cons x y) board))
\end{lstlisting}

Next we define \code{mark-hit}, which takes a player's board and
removes a piece if the guessed coordinate is present. We 
return a pair of the updated board and a boolean indicating
whether the guess was a hit:

\begin{lstlisting}[language=Racket,escapechar=|,name=example]
(define (mark-hit board x y)
  (if (null? board)
      (cons board false)
      (let* ([fst (car board)]
             [rst (cdr board)])
        (if (and (= (car fst) x)
                 (= (cdr fst) y))
            (cons rst true)
            (let ([rst+b (mark-hit rst x y)])
              (cons (cons fst
                          (car rst+b))
                    (cdr rst+b)))))))
\end{lstlisting}

Although \code{mark-hit} will operate on sensitive data (the game
boards), it is written without any special machinery to maintain the
secrecy of \code{board}. Protecting data {w.r.t.} policies is instead handled
automatically and implicitly by a runtime monitor.
When Alice and Bob want to play a game, they both create a
\emph{label} to protect their data. A label is unique id mapped to a policy predicate
that takes a key (e.g., the current user's name) and returns true or false to indicate
permission for the label. Alice's label is used to annotate the data she wants
to be kept secret. Supposing Alice chooses to be player 1, she may use the
following label:

\begin{lstlisting}[language=Racket,escapechar=|,name=example]
(define alice-label (let-label l (lambda (x) (= 1 x))) l)
\end{lstlisting}

This code illustrates label creation, policy predicates, and the first-class
nature of labels. The policy predicate $\texttt{($\lambda$ (\blue{x}) (= 1 \blue{x}))}$
grants permission to player 1 only and is associated with the dynamically generated
label l (returned and bound to \texttt{\blue{alice-label}}).
Bob would use a similar policy (but for player $2$ instead of $1$). At runtime,
the \code{let-label} form creates a label \fcol{$\ell_{A}$} and binds
it to a closure for its policy predicate. When Alice wants to protect a
value, she creates a facet annotated with her label and two
\emph{branches}. The positive (left) branch represents the value as it
should appear to her, and the negative (right) to everyone else:

\begin{lstlisting}[language=Racket,escapechar=|,name=example]
  (define alice-board
          (facet alice-label (add-pieces (makeboard) x|$_1$| y|$_1$| |$...$|) (|\bmth{\star}|)))
\end{lstlisting}

In the above example, \blue{$\star$} (lazy failure) is used in the negative branch
to ensure execution will fail if Bob tries to observe Alice's secret gameboard. To observe
Alice's gameboard, Bob can try to use $\code{(obs}\ \fcol{e_\ell}\ \blue{e_\text{key}}\ 
\blue{e_\text{fac}}\code{)}$ form, which takes a label, a key, and a faceted value.
Explicit observation projects a single label $\fcol{e_\ell}$ in faceted value $\blue{e_\text{fac}}$ 
to either its positive or negative facet, depending on whether the policy associated 
with $\fcol{e_\ell}$ returns true for key $\blue{e_\text{key}}$. If Bob tries to observe 
Alice's board, the policy predicate will return false (from \texttt{(= 1 $\blue{2}$)}) 
and Alice's negative facet $\blue{\star}$ will result.

In other applications, Alice may choose a sensible default value to
reveal to others---she may even want to create a \emph{nested}
facet. For example, a social-networking application may use a
nested facet consisting of two labels for
\rmth{\ell_{\text{Friends}}} and \rmth{\ell_{\text{Family}}}. A user
can then present three views of her social-media profile: \bmth{p_1} to
her family, containing her phone number and other contact information,
\bmth{p_2} to her friends showing her interests, and \bmth{p_3} to
everyone else, showing only her name and email.

\begingroup
\begin{gather*}
\pgfkeys{/pgf/inner sep=0pt} \pgfkeys{/pgf/inner xsep=0pt}
  \begin{forest}
    l sep=0,s sep=1em,
    for children={l sep=0,s sep=1em},
    [\bmth{\ell_{\text{Family}}} [\bmth{p_1}] [\bmth{\ell_{\text{Friends}}} [\bmth{p_2}] [\bmth{p_3}]]]
  \end{forest}
\end{gather*}
\endgroup

As a game of Battleship progresses, Alice and Bob make guesses in turn, and
driver code calls the function \code{mark-hit} with each of their respective
(faceted) game boards to record the attack. However, because Alice and Bob's
game boards are both faceted values, \code{mark-hit} cannot be directly applied as
in normal execution. Instead, faceted execution ``splits'' the evaluation of
the function application over both facets, running it first on the positive
branch, then again on the negative branch. Finally, the results of
each branch are combined again to produce a new faceted value. This allows FE
to avoid needing to reason about labels and policies until an explicit observation
point where a policy is checked and a faceted value is projected to one of its
facets.

%% \begin{center}
%%   \begin{tikzpicture}
%% 	  \node (0) at (0, 1) {\lstinline|(mark-hit |{\color{\colorMATH}\(\facet{\ell_A}{v^+}{v^-} ~~ x ~~y\)}\lstinline|)|};
%% 	  \node (1) at (2, 0)  {\lstinline|(mark-hit |{\color{\colorMATH}\(v^- ~~ x ~~ y\)}\lstinline|)|{\color{\colorMATH}\(~~= v''\)}};
%% 	  \node (2) at (-2, 0) {\lstinline|(mark-hit |{\color{\colorMATH}\(v^+ ~~ x ~~ y\)}\lstinline|)|{\color{\colorMATH}\(~~= v'\)}};
%% 	  \node (3) at (0, -1) {{\color{\colorMATH}\(\facet{\ell_A}{v'}{v''}\)}};
%% 	  \draw [->] (0) edge (1);
%% 	  \draw [->] (0) edge (2);
%% 	  \draw [->] (1) edge (3);
%% 	  \draw [->] (2) edge (3);
%%   \end{tikzpicture}
%% \end{center}

Because the applied function can be stateful, faceted execution also
maintains the current privilege level in a \emph{program counter} (PC). The
program counter is a property of the current evaluation context and is used
to build facets when writes are made to the store in a privileged context.
For example, if a stateful function ``splits'' when applied on both the positive and negative
facets of a value faceted by a label $\fcol{\ell}$, and on the positive branch
the function uses $\texttt{set!}$ to mutate a variable $x$ from $2$ to $3$,
FE semantics will set $x$ to $\facet{\ell}{3}{2}$ so that the value $3$
cannot be leaked from the secure context (speculative execution under the
\rmth{+\ell} facet). This is because, for the duration of the app ``split'',
the evaluation context records that all values are implicitly guarded by \rmth{+\ell}
and then \rmth{-\ell}, respectively. If the semantics for $\texttt{set!}$ does
not make this faceting explicit, a sensitive value can leak from one PC to another.
We expand upon these subtleties in Section~\ref{sec:semantics},
where we present a complete semantics for faceted execution.

After making various moves, we eventually want to reveal the game
boards, pulling the positive view out of \code{alice-board} to display
Alice's board. To do this, we must \emph{observe} the facet with an
\code{obs} form. Because Alice's board is faceted with
\code{alice-label}, we specify that we want to observe
\code{alice-label} and pass in an argument to that label showing that
Alice is indeed the person observing the facet:

\begin{lstlisting}[language=Racket,escapechar=|,name=example]
(obs alice-label 1 alice-board) ; Returns Alice's board
\end{lstlisting}

\section{A Formal Semantics for Faceted Execution}
\label{sec:semantics}

\begin{figure*}
\begin{displaymath}
  \begin{array}{lcrcl}
    \blue{c} & \blue{\in} & \blue{const} & \blue{::=} & \blue{'() \mid \code{true} \mid \code{false} \mid \ldots} \\
    \blue{x} & \blue{\in} & \blue{var} & \blue{::=} & \langle\textit{program variables}\rangle  \\
    \blue{e} & \blue{\in} & \blue{exp} & \blue{::=} & \blue{c} \mid \blue{x} \\
    & & & \blue{\mid} & \blue{(\lambda ~ (x) ~ e) \mid (e ~ e)} \\
    & & & \blue{\mid} & \blue{(\code{box} ~ e) \mid (\code{unbox} ~ e) \mid (\code{set!} ~ e ~ e)} \\
    & & & \blue{\mid} & \blue{(\code{let-label} ~ x ~ e ~ e)} \\
    & & & \blue{\mid} & \blue{(\code{facet} ~ e ~ e ~ e)} \\
    & & & \blue{\mid} & \blue{(\code{obs} ~ e ~ e ~ e)} \\
  \end{array}
\end{displaymath}
\caption{Syntax of \lang.}
\label{fig:syntax}
\end{figure*}

\begin{figure*}
    \begin{tabular}{cc}
      \begin{minipage}{.5\textwidth}
        \begin{displaymath}
          \begin{array}{rcrcl}
            \blue{\alpha} & \blue{\in} & \blue{\text{addr}} & \blue{=} & \blue{\ldots} \\
            \blue{bv} & \blue{\in} & \blue{\text{base-val}} & \blue{::=} & \blue{c \mid \alpha \mid \langle \lambda x.~ e, \rho \rangle \mid \star}\\
            \fcol{v} & \blue{\in} & \blue{\text{faceted-val}} & \blue{::=} & \blue{bv \mid} ~ \fcol{\facet{\blue{\alpha}}{v}{v}}
            \\
          \end{array}
        \end{displaymath}
      \end{minipage}
      &
      \begin{minipage}{.5\textwidth}
        \begin{displaymath}
          \begin{array}{rcrcl}
            \fcol{b} & \fcol{\in} & \fcol{\text{branch}} & \fcol{::=} & \fcol{+ \ell \mid - \ell} \\
            \fcol{pc} & \fcol{\in} & \fcol{\text{PC}} & \fcol{=} & \fcol{\wp(\text{branch})} \\
            \blue{\rho} & \blue{\in} & \blue{\text{env}} & \blue{=} & \blue{\text{var} \rightharpoonup v}\\
            \blue{\sigma} & \blue{\in} & \blue{\text{store}} & \blue{=} & \blue{\text{addr} \rightharpoonup v}
            \\
          \end{array}
        \end{displaymath}
      \end{minipage}
    \end{tabular}
    \\
    \hfill \textit{(Expression Evaluation)} \boxed{\blue{e, \rho, \sigma \Downarrow_{pc}^E \sigma, v}}
    \\
    \begin{displaymath}
      \begin{array}{c}
        \begingroup
        \color{colorMATH}
        \begin{array}{cccc}
          \inferrule[\textsc{Const}]{ }
                                    {c, \rho, \sigma \Downarrow_{pc}^E \sigma, c}
                    & 
          \inferrule[\textsc{Var}]{ }
                                  {x, \rho, \sigma \Downarrow_{pc}^E \sigma, \rho(x)}
                    & 
          \inferrule[\textsc{Lambda}]{ }
                                     {\lambda x. ~ e, \rho, \sigma \Downarrow_{pc}^E \sigma, \langle \lambda x. ~ e, \rho \rangle}
                    &
          \inferrule[\textsc{Apply}]{e_1, \rho, \sigma \Downarrow_{pc}^E \sigma', v_1 \and
                                     e_2, \rho, \sigma' \Downarrow_{pc}^E \sigma'', v_2 \\\\
                                     \fcol{(v_1 ~ v_2), \rho, \sigma'' \Downarrow_{pc}^A \sigma''', v'}}
                                    {(e_1 ~ e_2), \rho, \sigma \Downarrow_{pc}^E \sigma''', v'}
        \end{array}
       \endgroup
        \\\\
        \begingroup
        \color{colorMATH}
        \begin{array}{ccc}
          \inferrule[\textsc{Box}]
                    {e, \rho, \sigma \Downarrow_{pc}^E \sigma', v \and
                      \alpha \not\in dom(\sigma') \\\\
                      \sigma'' = \sigma' [\alpha \mapsto \formfacet{pc}{v}{\star}]}
                    {(\code{box} ~ e), \rho, \sigma \Downarrow_{pc}^E \sigma'', \alpha}
                    & 
          \inferrule[\textsc{Unbox}]
                    {e, \rho, \sigma \Downarrow_{pc}^E \sigma', v \\\\
                     v' = \fcol{\var{read}(\sigma',v,pc)}}
                    {(\code{unbox} ~e), \rho, \sigma \Downarrow_{pc}^E \sigma', v'}
                    & 
          \inferrule[\textsc{Set}]
                    {e_1, \rho, \sigma \Downarrow_{pc}^E \sigma', v_1 \and
                     e_2, \rho, \sigma' \Downarrow_{pc}^E \sigma'', v_2 \\\\
                     \sigma''' = \fcol{\var{write}(\sigma'', v_1, pc, v_2)}}
                    {(\code{set!} ~ e_1 ~ e_2), \rho, \sigma \Downarrow_{pc}^E \sigma''', v_2}
        \end{array}
       \endgroup
       \\\\
       \hfill \textit{(Facet Creation)}
       \\
       \begin{array}{ccc}
          {\color{colorFAC}
          \inferrule[\textsc{Fac-Create-Split}]
          {e_1, \rho, \sigma \Downarrow_{pc}^E \sigma', \ell \and
            \{+\ell, -\ell\} \cap pc = \varnothing \\\\
           e_2, \rho, \sigma' \Downarrow_{pc \cup \{+l\}}^E \sigma'', v_1 \and
           e_3, \rho, \sigma'' \Downarrow_{pc \cup \{-l\}}^E \sigma''', v_2 \\\\
           v = \formfacet{pc \cup \{+\ell\}}{v_1}{v_2}}
          {(\code{fac} ~ e_1 ~ e_2 ~ e_3), \rho, \sigma \Downarrow_{pc}^E \sigma''', v}}
          & 
          {\color{colorFAC}
          \inferrule[\textsc{Fac-Create-Pos}]
          {e_1, \rho, \sigma \Downarrow_{pc}^E \sigma', \ell \\\\
           + \ell \in pc \\\\
           e_2, \rho, \sigma' \Downarrow_{pc}^E \sigma'', v}
          {(\code{fac} ~ e_1 ~ e_2 ~ e_3), \rho, \sigma \Downarrow_{pc}^E \sigma'', v}}
          & 
          {\color{colorFAC}
          \inferrule[\textsc{Fac-Create-Neg}]
          {e_1, \rho, \sigma \Downarrow_{pc}^E \sigma', \ell \\\\
           - \ell \in pc \\\\
           e_3, \rho, \sigma' \Downarrow_{pc}^E \sigma'', v}
          {(\code{fac} ~ e_1 ~ e_2 ~ e_3), \rho, \sigma \Downarrow_{pc}^E \sigma'', v}}     \end{array}

      \end{array}
    \end{displaymath}
    \\
    \hfill \textit{(Label Creation / Observation)}
    \\
    \[
    \begin{array}{cc}
      {\color{colorFAC}
        \inferrule[\textsc{Let-Label}]
          {e_1, \rho, \sigma \Downarrow_{pc}^E \sigma', \langle \lambda x. ~ e, \rho' \rangle \\\\
           \alpha \not\in dom(\sigma') \and \sigma'' = \sigma' [\alpha \mapsto \langle \lambda x. ~ e, \rho' \rangle] \\\\
           e_2, \rho [ \ell \mapsto \alpha ], \sigma'' \Downarrow_{pc}^E \sigma''', v}
          {(\code{let-label} ~ \ell ~ e_1 ~ e_2), \rho, \sigma \Downarrow_{pc}^E \sigma''', v}}
          & 
          {\color{colorFAC}
          \inferrule[\textsc{Obs}]
                    {e_1, \rho, \sigma \Downarrow_{pc}^E \sigma', \ell
                      \and e_2, \rho, \sigma' \Downarrow_{pc}^E \sigma'', v  \\\\
                      \langle (\lambda x.~e), \rho' \rangle = \sigma''(\ell) \\\\
                      e, \rho' [x \mapsto v], \sigma'' \Downarrow_{pc}^E \sigma''', v^{\pm} \\\\
                      e_3, \rho, \sigma''' \Downarrow_{pc}^E \sigma'''', v' \\\\
                      v'' = \fcol{\var{obs}(\ell, v', v^{\pm})}}
                    {(\code{obs} ~ e_1 ~ e_2 ~ e_3), \rho, \sigma \Downarrow_{pc}^E \sigma'''', v''}}
    \end{array}
    \]
    \\
    \hfill \textit{(Possibly-Faceted Application)} \boxed{\fcol{(v_1 ~ v_2), \rho, \sigma \Downarrow_{pc}^A \sigma, v}}
    \\
    \begingroup
    \color{colorMATH}
    \begin{displaymath}
      \begin{array}{c}
        \begin{array}{ccc}
          {\color{colorMATH}
          \inferrule[\textsc{App-$\star$}]
          { }
          {(\star ~ v), \rho, \sigma \Downarrow_{pc}^A \sigma, \star}}
          & 
          {\color{colorMATH}
          \inferrule[\textsc{App-Base}]
          {e, \rho'[x \mapsto v], \sigma \Downarrow_{pc}^E \sigma', v' }
          {(\langle \lambda x.~e, \rho' \rangle ~ v), \rho, \sigma \Downarrow_{pc}^A \sigma', v'}}
          & 
          {\color{colorFAC}
          \inferrule[\textsc{App-Split}]
          {\{+\ell, -\ell\} \cap pc = \varnothing \and
            (v^+ ~ v), \rho, \sigma \Downarrow_{pc \cup \{+\ell\}}^A \sigma', {v^+}' \\\\
            (v^- ~ v), \rho, \sigma' \Downarrow_{pc \cup \{-\ell\}}^A \sigma'', {v^-}' \and
            v' = \formfacet{\{+ \ell\}}{{v^+}'}{{v^-}'}}
          {(\facet{\ell}{v^+}{v^-} ~ v), \rho, \sigma \Downarrow_{pc}^A \sigma'', v'}}
        \end{array}
        \\
        \begin{array}{cc}
        \\\\
          {\color{colorFAC}
          \inferrule[\textsc{App-Facet-Pos}]
          {+\ell \in pc \\\\
            (v^+ ~ v), \rho, \sigma \Downarrow_{pc}^A \sigma', v'}
          {(\facet{\ell}{v^+}{v^-} ~ v), \rho, \sigma \Downarrow_{pc}^A \sigma', v'}}
          & 
          {\color{colorFAC}
          \inferrule[\textsc{App-Facet-Neg}]
          {-\ell \in pc \\\\
            (v^- ~ v), \rho, \sigma \Downarrow_{pc}^A \sigma', v'}
          {(\facet{\ell}{v^+}{v^-} ~ v), \rho, \sigma \Downarrow_{pc}^A \sigma', v'}}
        \end{array}
      \end{array}
    \end{displaymath}
    \endgroup
    \caption{Semantics of Faceted Execution}
    \label{fig:semantics}
\end{figure*}

We now present a semantics for a core language (\lang) which includes
facets. Our presentation largely mirrors that of Austin et
al.~\cite{Austin:2013}. The syntax of our language---reminiscent of
Scheme---is shown in Figure~\ref{fig:syntax}. \lang extends the lambda
calculus with references (which have interactions with facets in a subtle
way) and three forms unique to faceted execution: facet construction, 
label creation, and facet observation.

Our semantics is shown in Figure~\ref{fig:semantics} \cite{Micinski:2019}.
As \lang is an
extension of the lambda calculus with references, we present the parts
unique to faceted execution in \fcol{red}, while keeping the lambda
calculus with references in \blue{blue}. Base values in our semantics
include addresses (used for boxes), constants, and closures. We also
include a kind of lazy failure (\bmth{\star}), which is necessary for
defining store update within a protected context.

Values in our semantics are either (unfaceted) base values or facets
composed of a label and two branches. Facets can nest, allowing trees
of faceted values. We use the term \emph{branches} to refer to
positive or negated labels. Collections of branches define the program
counter \rmth{pc}, which tracks the set of branches in the current
context. For example, to apply a faceted function to a value (as in
the application of $\facet{\ell}{\blue{\lambda x.~0}}{\blue{\lambda x.~1}}$),
the semantics first applies \bmth{\lambda x.~0} while extending \rmth{pc}
with \rmth{+\ell}, then applies the negative branch extending
\rmth{pc} with \rmth{-\ell}.

The reduction relation \bmth{e, \rho, \sigma \Downarrow_{pc}^E \sigma,
  v} reduces an expression, environment, and store to a resulting
store and value. The first three rules (all in blue) are unchanged
from the standard interpretation in the lambda calculus. The
\textsc{Apply} rule calls out to the helper relation \rmth{(v~v),
  \rho, \sigma \Downarrow_{pc}^A \sigma, v}, which applies a
possibly-faceted value to an argument: if the value being applied is a
plain (unfaceted) closure, the \textsc{App-Base} rule (in blue, as it
is unchanged from the lambda calculus) applies it and returns
immediately to the \bmth{\Downarrow_{pc}^E} relation.

In the case that a faceted value is applied, \rmth{\Downarrow_{pc}^A}
performs one of three functions, based on the relation of \rmth{\ell}
to \rmth{pc}. If there is no occurrence of either \rmth{+\ell} or
\rmth{-\ell} in \rmth{pc}, then the semantics has not yet branched on
\rmth{\ell}, and therefore must split the application. To do this, it
applies both the positive and negative branches after extending
\rmth{pc}. After reducing both branches to values, the results are
formed into a facet. If \rmth{+\ell \in pc}, then the semantics has
already branched on the label \rmth{\ell}, so splitting would be
redundant. In this case, \rmth{\Downarrow_{pc}^A} simply selects the
positive branch to apply and continues without splitting. The
symmetric case occurs in \textsc{App-Facet-Neg}. Facet formation
follows this pattern, accounting for the relation of \rmth{\ell} to
\rmth{pc}.

The rules \textsc{Box}, \textsc{Unbox}, and \textsc{Set} appear
similar to the standard implementation of boxes, but employ several
meta-functions to do their work. This is because box creation, reads,
and writes may occur within a privileged context, and care must be
taken to form facets when \rmth{pc} is nonempty. To understand why,
consider the following example\footnote{Our formal semantics elides
  \code{if}, though it may be obtained via a Church encoding if
  desired as in Austin et al.\cite{Austin:2012}. Our implementation
  includes \code{if} but not other constructs such as \code{cond}}:

\begin{lstlisting}[language=Racket,escapechar=|,name=boxexample]
(define x (box 0))
(if (= (facet alice 0 1) 0)
  (set! x 0)
  (set! x 1))
(unbox x)
\end{lstlisting}

If we do nothing special to account for the fact that the program
branches on the facet, control flow implicitly launders the value
through the box to an unfaceted value. To fix this, we form a facet by
taking into account \rmth{pc} and forming a facet using the
meta-function \rmth{\formfacet{\ell}{v^+}{v^d}}. This meta-function is
defined in Figure~\ref{fig:metafunctions}, and takes three arguments:
the current \rmth{pc}, a positive view (\rmth{v^+}), and a ``default''
view (\rmth{v^d}). Facet construction builds a facet with a spine
corresponding to all of the labels in \rmth{pc}, and inserts
\rmth{v^+} at the focus corresponding to \rmth{pc}, putting the
default value \rmth{v^d} along all other branches. In the \code{box}
form, the facet uses a default value of \bmth{\star}. In terms of our
above example, this means that along the positive branch \code{x}
would be set to \facet{\code{alice}}{1}{\star} (as the false branch of
the \code{if} is taken), and along the subsequent negative branch
\code{x} is extended to \facet{\code{alice}}{1}{0}.

\begin{figure*}
      %\begin{displaymath}
      %  \begin{array}{rclrr}
      %    {read(\sigma, \alpha, pc)} & {=} & {\sigma(\alpha)} & & \\
      %    {read(\sigma, \facet{\ell}{v_1}{v_2}), pc} & = & {read(\sigma, v_1, pc)} & \textit{if} ~ {+\ell} \in pc & \\
      %    {read(\sigma, \facet{\ell}{v_1}{v_2}), pc} & = & {read(\sigma, v_2, pc)} & \textit{if} ~ {-\ell} \in pc & \\
      %    {read(\sigma, \facet{\ell}{v_1}{v_2}), pc} & = & {\formfacet{\ell}{read(\sigma, v_1, pc)}{read(\sigma, v_2, pc)}} & \textit{otherwise} &
       % \end{array}
      %\end{displaymath}
      \begin{displaymath}
        \begin{array}{rclrr}
          {\formfacet{\varnothing}{v^+}{v^d}} & {=} & {v^+} \\
          {\formfacet{\{+\ell\} \cup rest}{v^+}{v^d}} & {=} & {\facet{\ell}{\formfacet{rest}{v^+}{v^d}}{v^d}} \\
          {\formfacet{\{-\ell\} \cup rest}{v^+}{v^d}} & {=} & {\facet{\ell}{v^d}{\formfacet{rest}{v^+}{v^d}}}
        \end{array}
      \end{displaymath}
      \\
      \begin{displaymath}
        \begin{array}{rclrr}
          {\var{write}(\sigma, \alpha, pc, v)} & {=} & {\sigma[\alpha ~ := \formfacet{pc}{v}{\sigma(\alpha)}]} & & \\
          {\var{write}(\sigma, \facet{\ell}{v_1}{v_2}), pc, v} & = & {\sigma''} & \textit{where} ~ {\sigma' = \var{write}(\sigma, v_1, pc \cup \{+\ell\}, v)} \\
          & & & \textit{and} ~ {\sigma'' = \var{write}(\sigma', v_2, pc \cup \{-\ell\}, v)}
        \end{array}
      \end{displaymath}
      %\\
      %\begin{displaymath}
      %  \begin{array}{rclrr}
      %    {obs(\ell, b, bv)} & {=} & {bv} & & \\
      %    {obs(\ell, true,} \facet{\ell}{v_1}{v_2}) & = & {v_1} & & \\
      %    {obs(\ell, false,} \facet{\ell}{v_1}{v_2}) & = & {v_2} & & \\
      %    {obs(\ell, b,} \facet{\ell'}{v_1}{v_2}) & = & \facet{\ell'}{obs(\ell,b,v_1)}{obs(\ell,b,v_2)} & \textit{where} & {\ell} \neq {\ell'}
      %  \end{array}
      %\end{displaymath}
      \caption{Meta-functions used in our semantics}
      \label{fig:metafunctions}
\end{figure*}

Label creation allocates a label as a fresh address in the store,
binding the specified label predicate and adding it to the
environment. Labels must be store-allocated rather than bound in the
lexical environment, as the latter would allow the label to be rebound
by anyone using the facet:

\begin{lstlisting}[language=Racket,escapechar=|,name=rebindingexample]
(define alice-label (let-label l (lambda (x) (= x alice)) l))
(define x (facet alice-label 1 0))
(let ([alice-label (let-label l (lambda (x) true) l)]) |\label{lineno:alicelab}|
  (obs alice-label 1 x)) ; Should return |\facet{\texttt{alice-label}}{1}{0}|
\end{lstlisting}

The \code{obs} form in the above example ought to return
\facet{\code{alice-label}}{1}{0}. But if we pull labels from the
lexical environment, the binding on line~\ref{lineno:alicelab} shadows
the policy originally associated with the facet.

Last, observation evaluates the label expression to an address and
executes the associated predicate. Once this is done, \textsc{Obs}
uses the \rmth{\var{obs}} meta-function to select the appropriate
branch based on the value returned by the predicate associated with
the label. This meta-function accounts for the fact that the label
being observed may appear arbitrarily deep in the facet (or not at
all). As our implementation of \rmth{\var{obs}} is unchanged from its
definition in~\cite{Austin:2012}, we elide it here.

\section{Faceted Execution as Macros}
\label{sec:implementation}

The semantics of faceted execution is an extension of the lambda
calculus, leading to a natural question: can we use Racket's macros~\cite{Kohlbecker:1986, Dybvig:1992} to
extend Racket to faceted execution? We will see that the answer is
yes, and the translation from the big-step rules is surprisingly
straightforward. This section of our paper describes the design of
\racets, a prototype implementation of faceted execution using Racket
macros. In Section~\ref{sec:conclusion} we remark upon current
directions scaling \racets to the whole of Racket.

\paragraph*{Choosing a Representation for Facets, Labels, and Program Counters}

In setting out to implement facets, we must first choose how we will
represent facets, labels, and program counters. We have chosen to
implement facets simply as Racket \lstinline|struct|s, containing a
label along with positive and negative branches:

\begin{lstlisting}[language=Racket,escapechar=|,name=racets]
(struct facet (labelname left right))
\end{lstlisting}

Next, we must choose a representation of labels. At first
consideration, it appears sensible to represent labels simply as
closures. After all, labels are simply used as predicates testing
whether or not to reveal a facet's positive or negative
branch. Therefore, we represent labels as a pair of a name and a policy:

\begin{lstlisting}[language=Racket,escapechar=|,name=racets]
(struct labelpair (name pol))
\end{lstlisting}

Now that we have defined labels, we can define branches, which are
positive or negative labels:

\begin{lstlisting}[language=Racket,escapechar=|,name=racets]
(struct pos (lab))
(struct neg (lab))
\end{lstlisting}

Similarly, program counters are sets of branches. However, we must
still ask how we will keep track of the ``current'' program
counter. Our implementation uses Racket's parameters, though other
mechanisms (such as continuation marks~\cite{clements2001modeling}, to which parameters
macro-expand) can also be used. \racets defines the parameter
\code{current-pc}, and updates it as computation progresses:

\begin{lstlisting}[language=Racket,escapechar=|,name=racets]
(define current-pc (make-parameter (set)))
\end{lstlisting}

\paragraph*{Facet Creation}

Facet creation appears as three separate rules in
Figure~\ref{fig:semantics}, which we recapitulate here in three
distinct colors for each case:

\[\small
\begin{array}{ccc}
  {\color{brown}
    \inferrule[\textsc{Fac-Create-Pos}]
              {e_1, \rho, \sigma \Downarrow_{pc}^E \sigma', \ell \and
                \boxed{+ \ell \in pc} \\\\
                e_2, \rho, \sigma \Downarrow_{pc}^E \sigma'', v}
              {(\code{fac} ~ e_1 ~ e_2 ~ e_3), \rho, \sigma \Downarrow_{pc}^E \sigma'', v}}
  & 
  {\color{gray}
    \inferrule[\textsc{Fac-Create-Neg}]
              {e_1, \rho, \sigma \Downarrow_{pc}^E \sigma', \ell \and
                \boxed{- \ell \in pc} \\\\
                e_3, \rho, \sigma \Downarrow_{pc}^E \sigma'', v}
              {(\code{fac} ~ e_1 ~ e_2 ~ e_3), \rho, \sigma \Downarrow_{pc}^E \sigma'', v}}
  &
  {\color{purple}
    \inferrule[\textsc{Fac-Create-Split}]
              {e_1, \rho, \sigma \Downarrow_{pc}^E \sigma', \ell \and
                \boxed{\{+\ell, -\ell\} \cap pc = \varnothing} \\\\
                e_2, \rho, \sigma' \Downarrow_{pc \cup \{+\ell\}}^E \sigma'', v_1 \\\\
                e_3, \rho, \sigma'' \Downarrow_{pc \cup \{-\ell\}}^E \sigma''', v_2 \and
                v = \formfacet{pc}{v_1}{v_2}}
              {(\code{fac} ~ e_1 ~ e_2 ~ e_3), \rho, \sigma \Downarrow_{pc}^E \sigma''', v}}
\end{array}
\]

Translating these rules to Racket involves observing that each one
will apply under one of three disjoint circumstances (each of them
boxed in the above rules): \bmth{+\ell \in pc}, \bmth{-\ell \in pc},
or else \bmth{\{+\ell,-\ell\} \cap pc = \varnothing}. This is a common
idiom in our faceted semantics, as we often want to select the
appropriate branch of a facet if its label already exists in
\bmth{pc}.

At first glance, it may not be obvious that we even \emph{need} a macro
for facet creation. But according to our semantics, the following
snippet should produce \code{true} if \bmth{+\ell \in pc}:

\begin{lstlisting}[language=Racket,escapeinside={<@}{@>},name=other,numbers=none]
(facet l true (error "this shouldn't get evaluated if <@\bmth{+\ell \in pc}@>"))
\end{lstlisting}

If we were to implement \code{fac} as a function, it would force
evaluation of the negative branch, inconsistent with the semantics of
\textsc{Fac-Create-Pos}. We can implement each of these conditions as
a Racket macro by considering whether \bmth{\ell \in pc}, as shown in
Figure~\ref{fig:fac-create}. Each color in the listing corresponds to
the analogous semantic rule. The implementation of
\textsc{Fac-Create-Pos} and \textsc{Fac-Create-Neg} is relatively
straightforward, but \textsc{Fac-Create-Split} extends \bmth{pc} for
each branch and subsequently forms a facet. This function implements
canonicalizing facet construction, and (as the implementation is a
transliteration of that in Figure 6 of Austin et
al.~\cite{Austin:2012}) we omit its definition here.

\begin{figure}
{\small
\begin{lstlisting}[language=Racket,escapeinside={<@}{@>},name=racets]
(define-syntax-rule (facet l e1 e2)
  (cond
\end{lstlisting}
\vspace{-.7\baselineskip}
{\color{brown}
\begin{lstlisting}[language=Racket,escapeinside={<@}{@>},name=racets,identifierstyle=\color{brown}]
    [(set-member? (current-pc) (pos (facet-labelname l))) e1]
\end{lstlisting}
}
\vspace{-.7\baselineskip}
{\color{gray}
\begin{lstlisting}[language=Racket,escapeinside={<@}{@>},name=racets,identifierstyle=\color{gray}]
    [(set-member? (current-pc) (neg (facet-labelname l))) e2]
\end{lstlisting}
}
\vspace{-.7\baselineskip}
{\small\color{purple}
\begin{lstlisting}[language=Racket,escapeinside={<@}{@>},name=racets,identifierstyle=\color{purple},keywordstyle=\color{purple}]
    [else
     (let ([left (parameterize
                     ([current-pc (set-add (current-pc)
                                           (pos (facet-labelname l)))]))]
           [right (parameterize
                      ([current-pc (set-add (current-pc)
                                            (neg (facet-labelname l)))]))])
           (mkfacet (set-union (set (pos (labelpair-name l)))
                                    (current-pc))
                    v1 v2))]))
\end{lstlisting}}
}
\caption{Facet creation as a macro}
\label{fig:fac-create}
\end{figure}

\paragraph*{Label Creation}

As we chose a representation of labels as pairs of symbols (the
label's name) and closures (the predicate corresponding to the label),
label creation is relatively straightforward from the semantics:

\begin{lstlisting}[language=Racket,escapeinside={<@}{@>},name=racets]
(define-syntax-rule (let-label l (lambda xs e) body)
  (let ([l (labelpair (gensym 'lab)
                      (lambda xs e))])
    body))
\end{lstlisting}

\paragraph*{Faceted Boxes, Writes, and Observations}

Our faceted semantics includes explicit \code{box} and \code{unbox}
forms. This differs from Racket's semantics, where any variable may be
treated as a box due to assignment conversion. We have two main
options:

\begin{itemize}
\item Introduce an explicit \code{unbox} form in \racets, trusting the
  programmer to explicitly use our implementation of \code{unbox} on
  potentially-faceted objects.

\item Walk over Racket code (after macro-expansion via
  \code{local-expand}) transforming variable references to use
  explicit \code{unbox} forms from \racets.
\end{itemize}

For our prototype of \racets, we have chosen to implement the
first. This leads to a relatively simple implementation, but
essentially trusts the programmer to use \racets' \code{unbox} forms
when necessary.

\begin{figure}
\small
\begin{lstlisting}[language=Racket,escapeinside={<@}{@>},name=racets]
(define-syntax (ref-set! stx)
  (syntax-case stx ()
    [(_ var e)
     #`(let ([v e])
         (let write ([var var]
                     [pc (current-pc)])
           (if (box? var)
               ; <@$\color{colorFAC}\var{write}(\sigma, \alpha, pc, v)$@>
               (set-box! var (construct-facet (current-pc) v (unbox var)))
               ; Else split
               (mkfacet
                 (facet-labelname (unbox var))
                 ; <@$\color{colorFAC}\var{write}(\sigma, \alpha, pc \cup \{+l\}, v)$@>
                 (write
                   (facet-left (unbox var))
                   (set-add pc (pos (facet-labelname var))))
                 ; <@$\color{colorFAC}\var{write}(\sigma, \alpha, pc \cup \{-l\}, v)$@>
                 (write
                   (facet-right (unbox var))
                   (set-add pc (neg (facet-labelname var))))))))]))
\end{lstlisting}
\caption{\racets' implementation of \code{set!}}
\label{fig:setbang}
\end{figure}

\racets defines a \code{box} macro, along with \code{unbox} and
\code{set!}. We include the definition of \code{set!} in
Figure~\ref{fig:setbang}, which inlines the definition of the
\textit{write} metafunction from Section~\ref{sec:semantics} to
consider the case under which a facet is used when an address is
expected.

Facet observation is handled similarly, first evaluating the label to
produce a policy predicate, followed by evaluating the policy's
argument and a possibly-faceted value to observe. After applying the
policy its argument, we produce the value \bmth{v^{\pm}} and descend
down the facet until reaching either a base value or finding the
selected label (at which point we select the appropriate branch).

\paragraph*{Faceted Function Application}

By now we can anticipate a predictable pattern for implementing
faceted execution: check \bmth{pc} to decide whether to branch left,
right, or split. This is largely our strategy for handling faceted
function application, with a small twist: we need to be able to apply
functions from outside of \racets. For example, if we want to apply
builtin functions such as \code{display}, we need to be mindful of the
fact that these functions cannot work with faceted arguments.

To handle this, we implement a macro for the \code{lambda} form, to
tag \racets closures specifically (so that they are differentiated from
functions outside of the current module). In the case that a foreign
function is applied to a faceted value, our implementation of function
application wraps the function to be able to handle facets by
distributing the function through each branch of the facet.

In general it is unsafe to apply an unknown function to a faceted
value. This is because the unknown function may leak the facet's
private information as a side-effect. Therefore, our current
implementation of \racets allows programmers to apply external
functions, but does not make any guarantee of safety. A better
strategy may be to perform an \code{obs} before each call to a
potentially-unsafe function. In general, we believe module
interactions are a challenging problem in faceted execution, and we
leave its study to future work.

\section{Implementation and Evaluation}

We have implemented \racets as a set of Racket macros which can be
employed as a language using Racket's \texttt{\#lang reader}
facility. So far, we have included macros for many of Racket's core
forms including application, \code{if}, \code{lambda}, and
references. We leave others (including continuation marks) to future
work. Additionally, \racets does not support first-class control. There
has been recent work in handling exceptions in the context of faceted
execution~\cite{Austin:17}, however reconciling first-class control in
general remains (to our knowledge) an open problem.

Because of Racket's flexible macro system, our implementation of
faceted execution is much smaller than other systems: our core macros
comprise roughly 170 source lines of Racket, with another 120 lines of
library code to perform various facet-related functions.

Our implementation is currently available on Github at
\blue{\url{https://github.com/fordsec/racets}}

\paragraph*{Case Study: Battleship in Racets}

\begin{figure*}
  \subfloat[]{%
    \includegraphics[height=1.1in]{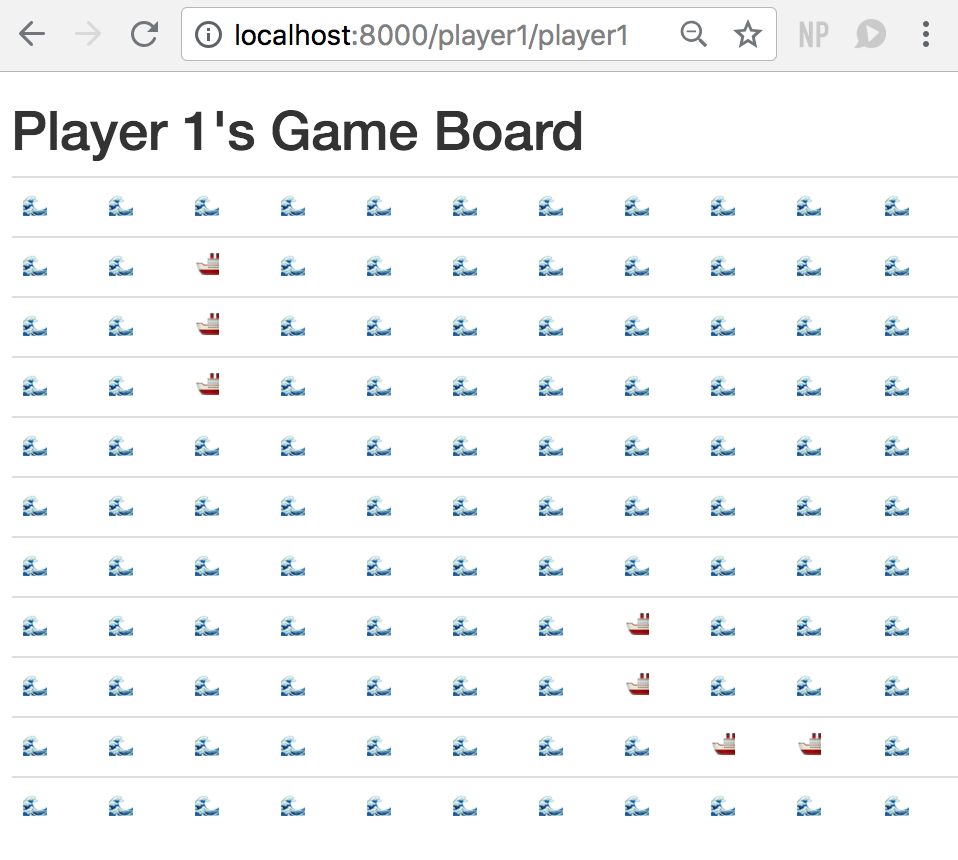}%
    \label{fig:player1}%
  } 
  \quad
  \subfloat[]{%
    \includegraphics[height=1.1in]{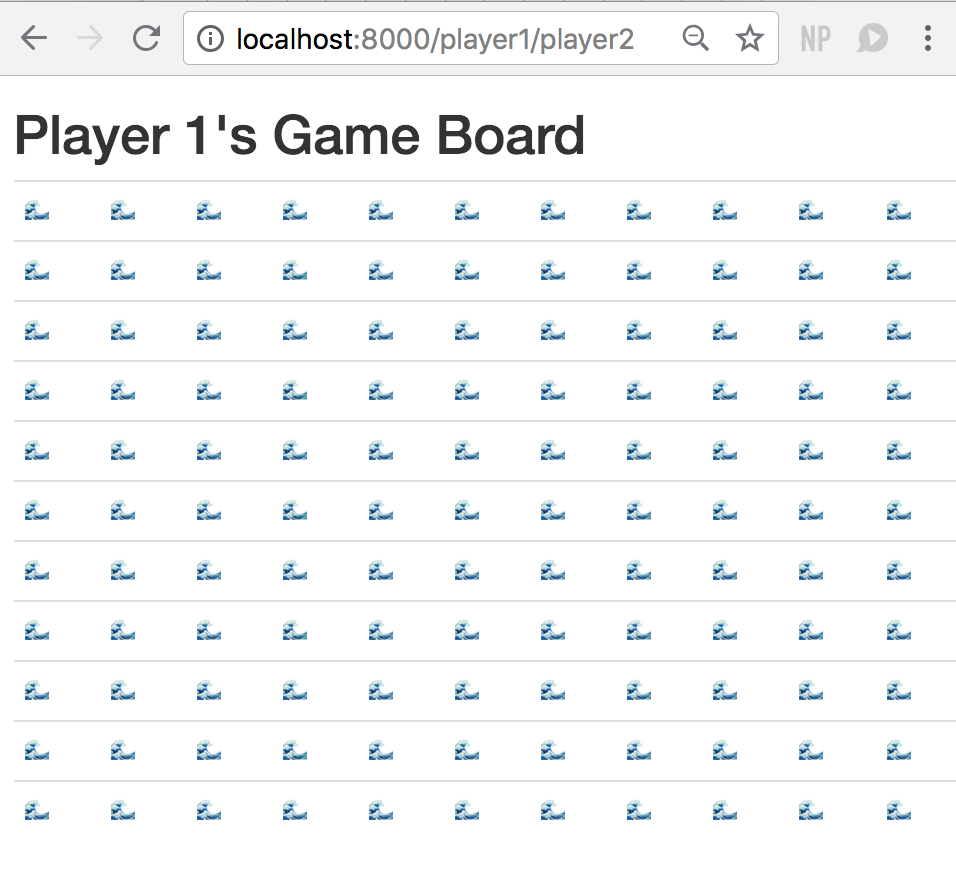}%
    \label{fig:player1player2}%
  }
  \quad
  \subfloat[]{%
    \includegraphics[height=.7in]{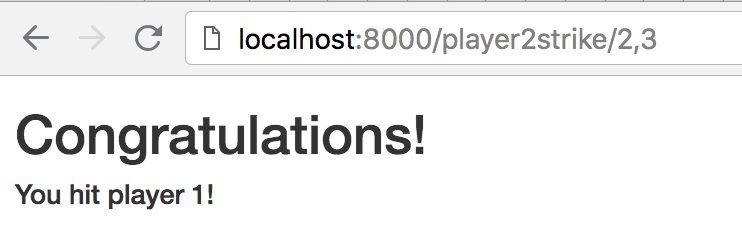}%
    \label{fig:player2strike}%
  }
  \vspace{.3in}
  \quad
  \subfloat[]{%
    \includegraphics[height=1.1in]{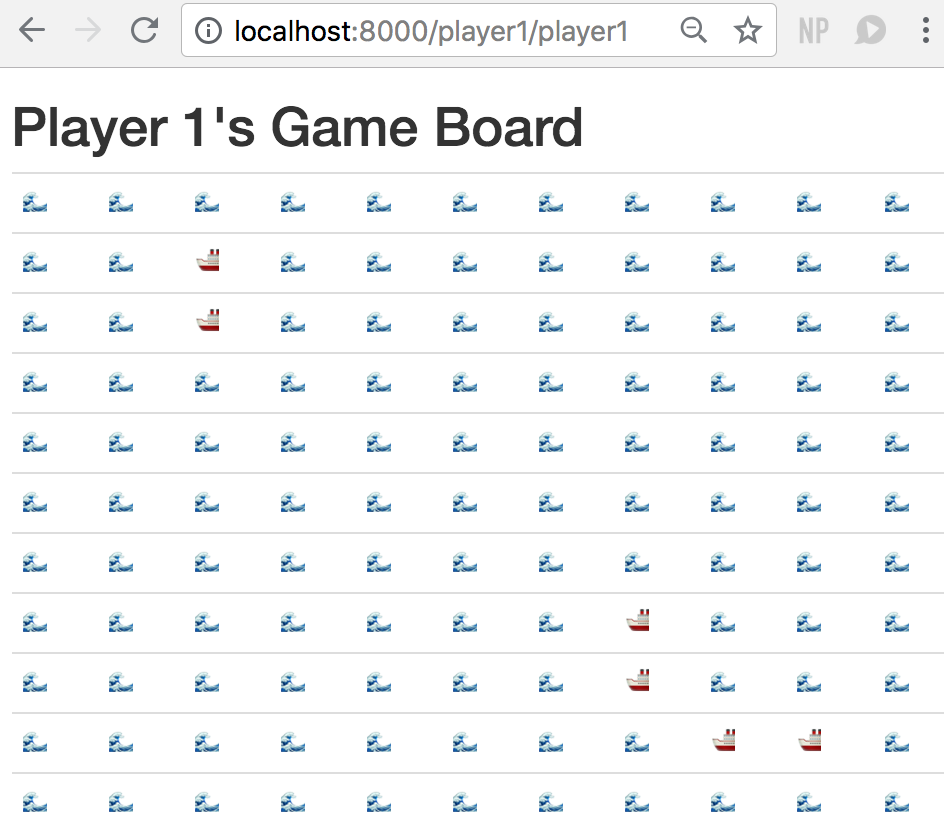}%
    \label{fig:player1afterstrike}%
  }
  \caption{Screenshots from our Battleship case study.}
  \label{fig:case}
\end{figure*}

To gain perspective on how Racets enables policy-agnostic programming,
we scaled our implementation from Section~\ref{sec:overview} to a
web-based game of Battleship written in Racets. Our implementation
uses Racket's \texttt{web-server} framework~\cite{webserver}, which
defines an API for writing HTTP-based server applications. 

Figure~\ref{fig:case} shows several screenshots of our Battleship
application. The first screenshot~\ref{fig:player1} shows the board as
viewed by player 1 (using the route \texttt{/player1/player1}),
while the second shows the empty board observed when player 2 attempts
to view player 1's board. The screenshot in~\ref{fig:player2strike}
shows the response seen by player 2 upon a successful hit. Finally
screenshot~\ref{fig:player1afterstrike} shows player one's board with
the ship on \bmth{(2,3)} removed.

When running, our game server provides several routes that a user can
access:

\begin{lstlisting}[language=Racket,escapechar=|,name=case]
(define-values (dispatch generate-url)
  (dispatch-rules
    [("player1" (string-arg)) player1board]
    [("player2" (string-arg)) player2board]
    [("player1strike" (string-arg)) p1strike]
    [("player2strike" (string-arg)) p2strike]))
\end{lstlisting}

The route \texttt{/player1/<id>} (or \texttt{/player2/<id>}) renders
player 1's game board when viewed as \texttt{<id>}. We facet game
boards with policies that reveal player 1's board when \texttt{<id>}
is \texttt{player1}, and do the same with player 2:

\begin{lstlisting}[language=Racket,escapechar=|,name=case]
(define p1l (mkpol "player1"))
(define p1board
  (box (facet p1l
            (add-pieces (makeboard) '(1 2 ldots))
            (makeboard))))
\end{lstlisting}

The board is explicitly made into a box: this is because the board's
state will change as player 2 makes moves and eliminates pieces from
their board. The \code{player1board} function implements the logic to
render player 1's board as an HTML table. The argument \code{viewer}
corresponds to the \texttt{<id>} route argument, and is passed to
\code{player1board} by the framework. This argument is then used to
observe the board game

\begin{lstlisting}[language=Racket,escapechar=|,name=case]
(define player1board
  (ext-lambda (request viewer)
         (http-response "<h1>Player 1's Game Board</h1>"
                        (pretty-print 
                           (obs p1l name (deref p1board))))))
\end{lstlisting}

The implementation of \code{player1board} uses a special form
\code{ext-lambda}, discussed at the end of this section, to allow code
from Racets to be executed by the framework (which is not prepared to
execute faceted code).

The function \code{p1strike} allows player 1 to make a guess as to the
position of ships on player 2's board. The function parses the
position into two coordinates and then calls \code{mark-hit} to
perform the hit, updating player 2's board and then observing the
result to answer (to player 1) whether the result was a hit or not:

\begin{lstlisting}[language=Racket,escapechar=|,name=case]
(define p1strike
  (ext-lambda
   (request position)
   (let* ([x (char-to-num (string-ref position 0))]
          [y (char-to-num (string-ref position 2))]
          [ans (mark-hit p2board x y)]
      (set! p2board (car ans))
      (http-response
       (if (cdr (obs p2l "player2" ans))
           "<h1>Congratulations!</h1> <h4>You hit player 2!</h4>"
           "<p>No hit :(</p>")))))))
\end{lstlisting}

Note that we need to use an explicit \code{obs} form on
line 25. This is because---as \code{p2board} is
a facet---the result will also be a faceted value. When we want to
display the output to player 1, our code needs to explicitly observe
the answer, as \code{http-response} cannot accept a faceted value.

\paragraph*{Module Interactions in \racets}

There is a wealth of existing Racket code we may like to incorporate
into \racets programs. For example, our case study uses the
\texttt{web-server} framework for building web applications. However,
in general, we believe that interacting with code not written using
faceted execution is a challenging open problem, and we do not know of any
principled solutions in the literature.

One immediate problem in \racets is how to pass functions from \racets
to plain Racket code. For example, the \texttt{web-server} framework
is written in Racket, and does not know how to call tagged closures
from \racets. As a stopgap, we added an \code{ext-lambda} form to
\racets. This form allows creating a Racket-style lambda in Racets
that will be used by functions in other modules, necessary for the
implementation of our case study.

We plan to explore interactions with unfaceted code more in the
future, and believe it will an exciting direction. For example, once
execution escapes \racets, we have no guarantee that the privacy
policy won't be violated. One solution may be to implicitly perform an
\code{obs} based on the current \rmth{pc} at points where \racets
interacts with unfaceted modules. But we do not fully understand the
ramifications or ergonomics of this choice, and suspect there may be a
wide array of design choices to handle these module interactions
including security type systems and blame (to track which module
violated the privacy policy).

\section{Related Work}
\label{sec:related}

To the best of our knowledge, we are the first authors to present an
implementation of faceted execution using hygienic macros. There are
several threads of related work in dynamic information flow and
programming paradigms for information flow.

Information-flow was first formalized by \citet{Denning:1976}.
In her seminal work on a lattice model
for information flow, she outlined challenges and potential solutions
to static information-flow checking. Subsequently, \citet{Goguen:82}
defined noninterference, which formalized the idea that privileged
data should not influence publicly observable
outputs. \citet{Clarkson:08} later recognized that
information-flow properties fit into a class of program properties
that could not be characterized by a single trace of a program, but
rather a set of traces, and called these hyperproperties.

Along with definitions of information flow, there has also been
significant interest in mechanisms for enforcing information
flow. This work can be broadly divided into static and dynamic
enforcement mechanisms for information flow security. Of the
mechanisms for static information flow, security type systems have
gained the most use. First introduced by \citet{Volpano:1997}, these
type systems augment the binding environment to track the privilege of
variables and prevent writes to variables that would violate
noninterference. Myers leveraged this idea to produce Jif, a variant
of Java with an information-flow type
system~\cite{Myers:1999}. Security type systems have been subsequently
extended to accommodate concurrent programs~\cite{Zdancewic:03} and
flow sensitivity~\cite{Hunt:2006}. Faceted execution does not require
annotating the program with security types, but at the expense of
losing a static characterization of the program's security in its type
system.

\citet{Devriese:10} first introduced secure multi-execution as a
dynamic enforcement technique for information flow. Secure
multi-execution runs $2^k$ copies of a program in parallel, where each
run represents a subset of $\wp(\var{Prin})$, where $\var{Prin}$ is a
set of principals. For example, if the principals in the program are
Alice and Bob, secure multi-execution executes four copies of the
program: one that replaces all secret inputs by $\bot$, one that
replaces Bob's input by $\bot$ but Alice's input by the true input,
one for Bob's input, and one with access to all privilaged
information. When external effects are made (e.g., writing to disc),
the runtime can select which variant to use based on a policy. Secure
multi-execution prevents information flow violations at runtime by
ensuring that observations which violate the information-flow policy
receive a view of the data computed without access to the secret
inputs. Secure multi-execution has been extended in a variety of ways,
e.g., scaling to its implementation in web browsers~\cite{Bielova:11},
adding declassification in a granular way~\cite{Rafnsson:13}, and even
preventing side-channel attacks~\cite{Kashyap:11}.

As the number of principals increases, secure multi-execution's
overhead increases exponentially, unnecessarily duplicating work not
influenced by secret inputs. Austin et al. introduced faceted
execution as an optimization of secure multi-execution
in~\cite{Austin:2012}. Instead of treating the whole program as a
potentially-secret computation, faceted execution realizes that
influence can be tracked and propagated in a granular way using
facets. Notably, Austin et al.'s work does not include first-class
labels, as it was simulating secure multi-execution, where the
principals could not be dynamically generated.

At the same time, Yang et al. first implemented Jeeves, a language
allowing policy-agnostic programming~\cite{Yang:2012}. Policy-agnostic
programming takes the view that programs should be written without
regard to a particular privacy policy, because as the policy changes,
correctly updating program logic is cumbersome and
error-prone. Policy-agnostic programming was first implemented in the
domain-specific language Jeeves, using an SMT solver to decide which
view of secret data to reveal based on a policy. Later, both authors
collaborated to implement Jeeves using faceted
execution.~\cite{Austin:2013}. This formulation includes first-class
labels, and is the basis for our concrete semantics.

Several other efforts into dynamic analysis for information flow are
worth noting. \citet{Stefan:2011} first presented \code{LIO}---a monad
(with implementation in Haskell) that tracks privilege of the current
program counter and forbids effects that would violate the security
policy. It may be surprising that \code{LIO}
works well for Haskell programs, given that faceted execution is
more precise than \code{LIO}---allowing values to become faceted
rather than halting the program. One key difference is that Haskell
programs emphasize purity while languages such as
JavaScript (the original target of faceted execution) does not, so much
of the machinery for faceted execution's effect on the store is less
interesting. Several authors have implemented related systems to
\code{LIO}, including variants of faceted execution \cite{Schmitz:2016}
and variants of \code{LIO} that extend its
power to arbitrary monad transformers~\cite{james}. We believe that it
would be possible to implement a variant of our technique that would
give similar insights to programs using \code{LIO}, though much of the
interesting machinery for handling state may be unnecessary.

\section{Conclusion and Future Work}
\label{sec:conclusion}

In this paper, we have reviewed the operation of faceted execution, a linguistic paradigm
enabling policy-agnostic programming, and showed how it may be implemented within
the Racket programming system as a library of macros. As Racket macros permit core language
forms (including function application, $\lambda$-abstraction, conditionals, mutation, etc.)
to be rewritten arbitrarily, it is possible to modify the meaning of these forms to support
a faceted semantics directly. We call our prototype system \racets: Racket with Facets.

The advantage of this approach is that faceted, policy-agnostic, programs may be written
directly in Racket, making use of the wealth of Racket code already available.
A central challenge of this then, is how to ensure there is a sound (w.r.t. secure multi-execution)
and practical inter-operation between \racets and standard Racket (or other languages written as
libraries in Racket). Our approach to this has been to use a tagging scheme that identifies values
from \racets so untagged values may be treated by \racets as originating from a non-\racets language.
For example, a pure Racket function that is not tagged, being applied at a $\blue{\code{ppapp}}$
form in \racets, can be automatically lifted to support FE (so that it can split when applied
on a faceted value).

Our hypothesis is that this tagging scheme is key to permitting inter-operation
between \racets and Racket, and we have implemented a faceted, web-based game of Battleship
to explore this idea. We suspect that a more thorough investigation of likely idioms
for faceted, non-faceted module interaction is needed and that purely functional code
plays a special role as a degenerate case where arbitrary non-faceted code may be lifted
to operate over facted values without potential unsoundness. In the future, we plan to explore
these design choices in a more principled way and also to scale a static analysis \cite{Micinski:2019}
of faceted execution to fully expanded Racket so it may be applied directly to \racets.

\bibliographystyle{ACM-Reference-Format}
\bibliography{paper}
\end{document}